\documentclass[prb,twocolumn,showpacs,preprintnumbers,amsmath,amssymb]{revtex4}
\usepackage{graphicx}
\usepackage{dcolumn}
\usepackage{bm}

\DeclareMathOperator{\sgn}{sgn}

\begin{document}

\title{Carbon Nanotubes in Helically Modulated Potentials}
\author{P. J. Michalski}
\author{E. J. Mele}

\affiliation{Department of Physics and Astronomy, University of
Pennsylvania, Philadelphia, PA 19104}

\date{\today}

\begin{abstract}
We calculate effects of an applied helically symmetric potential on
the low energy electronic spectrum of a carbon nanotube in the
continuum approximation. The spectrum depends on the strength of
this potential and on a dimensionless geometrical parameter, $P$,
which is the ratio of the circumference of the nanotube to the pitch
of the helix. We find that the minimum band gap of a semiconducting
nanotube is reduced by an arbitrarily weak helical potential, and
for a given field strength there is an optimal $P$ which produces
the biggest change in the band gap. For metallic nanotubes the Fermi
velocity is reduced by this potential and for strong fields two
small gaps appear at the Fermi surface in addition to the gapless
Dirac point. A simple model is developed to estimate the magnitude
of the field strength and its effect on DNA-CNT complexes in an
aqueous solution. We find that under typical experimental conditions
the predicted effects of a helical potential are likely to be small
and we discuss several methods for increasing the size of these
effects.
\end{abstract}



\maketitle

\section{\label{sec1} Introduction}

In recent years it has become a common practice to functionalize
CNTs with water soluble, high molecular weight, quasi-linear
molecules such as synthetic polymers and
DNA.\cite{obe2001,zjs2003,msh2003} These molecules bind to
individual CNTs via van der Waals forces and form robust and stable
complexes with the nanotube.\cite{gk2004} The resulting complex is
easily dispersed in an aqueous solution because the hydrophobic
nanotube is screened from the water by the wrapping molecule, while
the hydrophilic regions of the DNA or polymer are free to interact
with the solvent. Forming such complexes between CNTs and
quasi-linear molecules is the only known way to disperse nanotubes
in aqueous media without using surfactants or chemically modifying
the CNT. Several practical applications have been realized as a
result of this technique.  DNA wrapped CNTs can be sorted by
diameter using density gradient ultra-centrifugation, and the
resulting mixture filtered to obtain a solution comprised almost
entirely of one nanotube species.\cite{ash2005}  DNA wrapped CNTs
are especially well suited for biological applications, such as
cellular markers,\cite{hbe2005} which are not possible with
surfactant dispersed CNTs because surfactant molecules generally
destroy biological systems.

Simulations and experimental observations indicate that polymers and
DNA can wrap around the exterior of a CNT in an ordered, helical
fashion.\cite{obe2001, zjs2003, zjss2003, ssh2002, gsb2006} In this
paper we will study the effect of a helical potential on the single
particle energy spectrum of a CNT. From a practical perspective,
such a calculation is needed because of the large number of
experiments performed on DNA-CNT complexes and changes in the band
structure induced by the helical potential must be taken into
account when interpreting experimental data. Also, while many
polymers easily dissociate from the nanotube upon a change in
solvent,\cite{obe2001} DNA is much more difficult to
remove.\cite{agh2006} For device applications, it will be useful to
know if a DNA-CNT complex can be substituted for a pristine CNT
without a change in response or loss of function.  From a more
fundamental perspective, our work provides scaling relationships
relating changes in the band structure to structural parameters such
as the nanotube radius and the pitch of the helix.

Two related studies have recently been published. In
Ref.~(\onlinecite{wf2006}), the authors modeled the nanotube as a
free electron gas confined to the surface of a cylinder and governed
by the non-relativistic Schrodinger equation. A helical potential
was introduced as a series of delta functions, similar to the
Kronig-Penney model. This study concluded that the total electronic
energy is an oscillatory function of the pitch, with several local
minima indicating preferred wrapping angles. However, it has been
shown that the low energy electronic structure of a CNT is better
described by the Dirac Hamiltonian for a massless relativistic
particle.\cite{aa1993,mk1997} While the results of
Ref.~(\onlinecite{wf2006}) may describe the effect of a helical
potential on a semiconducting nanowire, it does not generally
describe the electronic physics of a carbon nanotube. Although we do
not study the total electronic energy, our results below, obtained
within the long wavelength Dirac theory, do not show any oscillatory
behavior as a function of the pitch.

In Ref.~(\onlinecite{prot2007}), the authors studied the electronic
response of an armchair nanotube to an applied helical potential.
Only potentials commensurate with the nanotube lattice were
considered, and the high symmetry of the underlying armchair lattice
was important in facilitating calculations.  This study concluded
that the external potential opened up small band gaps in the
originally metallic nanotube.  To discover such an effect it is
important to consider the nanotube lattice; as such, the continuum
theory we develop below does not reproduce these tiny band gaps.  To
study a general chiral nanotube with an arbitrary helical potential
using the method of Ref.~(\onlinecite{prot2007}) would be a
formidable task, whereas the problem is accessible within the
continuum theory.  The tradeoff is that we miss higher order effects
such as tiny band gaps in otherwise metallic nanotubes.  In general
the effects missed by the continuum theory are so small as to be
unimportant.  The corrections introduced by higher order
considerations are discussed in the conclusion and in the two
appendices.

For semi-conducting tubes, we find that the band gap always closes
under an applied, arbitrarily weak helical potential.  For a given
polymer-CNT system, the change in the band gap is a relatively
sharply peaked function of pitch, with an optimum pitch maximizing
the response to the potential. As the pitch goes to zero, the effect
of the helical potential vanishes.  As the pitch goes to infinity,
there are two different behaviors depending on the strength of the
applied field: for weak fields the effect of the chiral potential
vanishes, for strong fields the band gap remains closed.  For
metallic tubes, we find that the Fermi velocity is a slowly
decreasing function of pitch.  As with semi-conducting tubes, as the
pitch goes to zero the effect of helical potential vanishes. As the
pitch goes to infinity, there are again two different behaviors
depending on the field strength. For strong fields two small gaps,
inversely proportional to the pitch, appear near the Fermi surface
in addition to the gapless Fermi point. For weak fields no such gaps
appear in the spectrum. In both cases the Fermi velocity is reduced
by the helical potential.

Early on it was conjectured that the underlying chirality of the
CNT might determine the pitch of an adsorbed polymer
 \cite{cga2001}.  However, it is now generally believed that the
adsorbed  species need not conform to the lattice structure of the
nanotube. The structure of most wrapping molecules is actually
incommensurate with the nanotube lattice, and deforming the
molecule to match the lattice can be energetically
unfavorable.\cite{cf2004} Simulations show that, depending on the
nucleotide base sequence, single stranded DNA (ssDNA) can wrap
around a given nanotube in a left-handed or right-handed helix, or
even bond linearly along the tube.\cite{zjs2003, gk2004}
Experiments show that identical strands of DNA will wrap with the
same handedness around enantiomeric pairs of
nanotubes.\cite{dbd2006} Additionally, recent simulations
demonstrate that DNA bases can adhere to a nanotube in hundreds of
stable configurations.\cite{mmp2007} All of this is strong
evidence that the underlying lattice of the nanotube plays little
to no role in determining the structure of the CNT-DNA (-polymer)
complex. The evidence suggests that the DNA (polymer) wraps in a
manner determined by its own physical properties (chemical
composition, elastic stiffness, size, etc.) and the CNT
radius.\cite{gsb2006} Furthermore, the interaction between the
nanotube and the wrapping molecule should be adequately described
by a position independent binding energy.\cite{cf2004} These
observations motivate us to ignore the discrete atomic nature of
the nanotube and treat it as a cylinder in the continuum limit in
our calculations below.

The remainder of this paper is organized as follows.  In
Sec.~\ref{sec2} we will describe our model and a convenient
coordinate transformation.  In Sec.~\ref{sec3} we will give analytic
and numerical solutions of the low energy spectrum for both metallic
and semi-conducting tubes. In Sec.~\ref{sec4} we use a simple model
of the wrapping molecule to estimate the magnitude of the applied
field and the resulting change in the nanotube band gap.  In
Sec.~\ref{sec5} we give a brief conclusion.

\section{\label{sec2} Model System}

As discussed in the introduction, the nanotube will be modeled as a
continuous cylinder subjected to an external helical potential. By
developing a long wavelength continuum theory we can avoid
complications that arise when considering electrons subjected to two
incommensurate potentials on the scale of the lattice
constant.\cite{dhof1976} The low energy electronic states of the
nanotube are obtained by expanding the graphene Hamiltonian around
the $\mathbf{K}$ and $\mathbf{K}^\prime$ points and applying
appropriate boundary conditions.\cite{aa1993,mk1997} In what follows
we consider the solutions near the $\mathbf{K}$ point since the
response to a static helical potential must be the same at
$\mathbf{K}$ and $\mathbf{K}^\prime$ due to time reversal symmetry.

A long wavelength continuum theory is appropriate if the pitch of
the potential is large compared to the nanotube lattice spacing.
Below we will investigate the electronic response of the nanotube as
a function of the pitch of the potential and the nanotube
circumference.  The important parameter will be the ratio of
circumference to pitch, and we will study the response of the
nanotube in the limits that this ratio goes to zero and as it goes
to infinity.  In the first limit we imagine fixing the nanotube
radius and allowing the pitch to go to infinity, where our long
wavelength description is certainly valid.  In the second limit the
long wavelength description is only valid if we imagine that the
pitch is fixed and the radius is allowed to go to infinity.  In the
limit where the radius is fixed and the pitch goes to zero the
underlying lattice structure becomes important and a continuum
theory is no longer appropriate.

\begin{figure}
\includegraphics[width=80 mm]{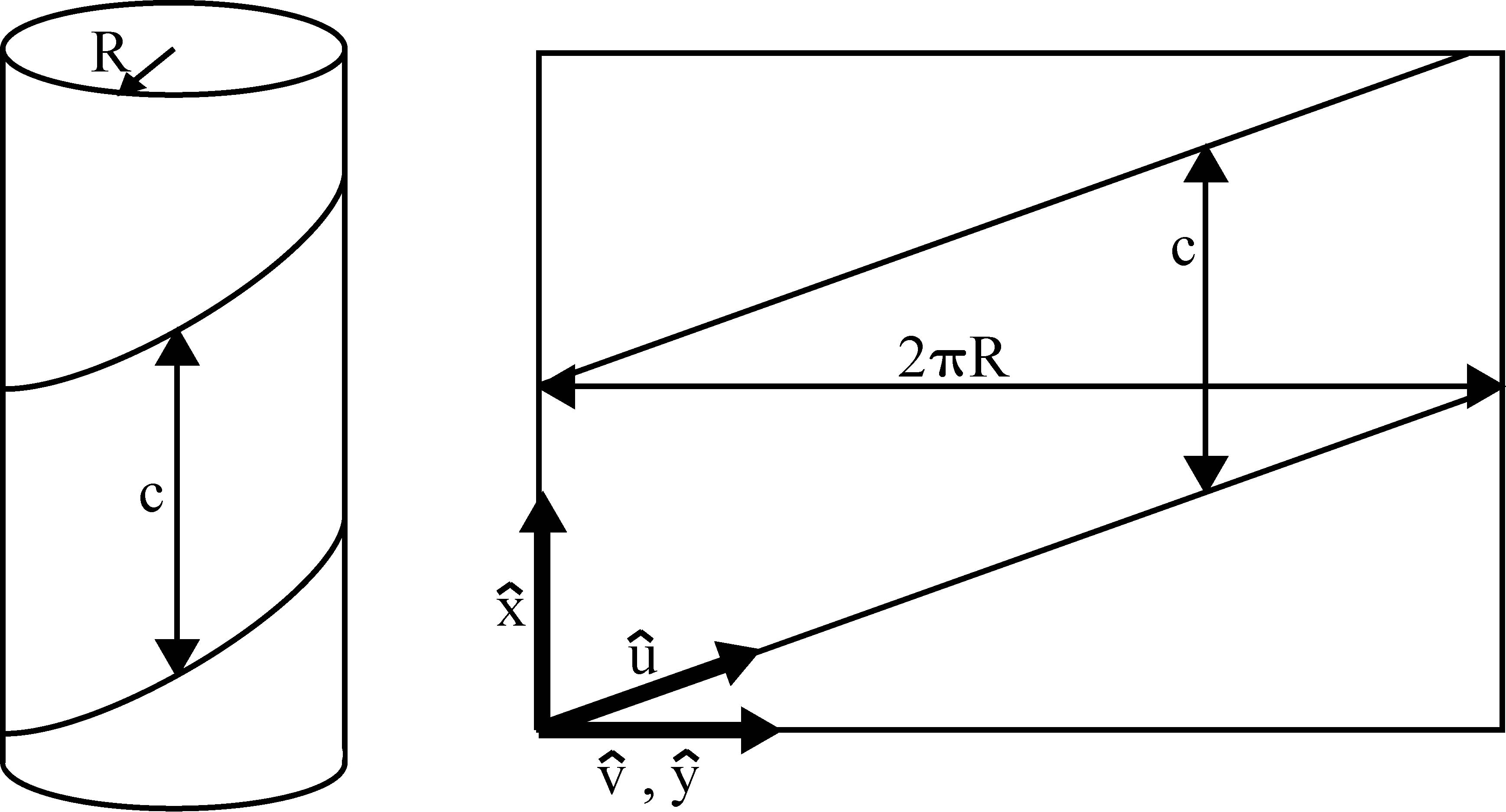}
\caption{\label{geometry} The tube on the left illustrates the
geometry of our model system.  Here, $R$ is the radius of the
nanotube and $c$ is the pitch of the helical potential.  On the
right is an ``unwrapped'' view of the nanotube surface, with the
pitch and the nanotube circumference, $2\pi R$, marked.  Also on the
right are the four relevant spatial unit vectors for this system,
the usual $\hat{x}$ and $\hat{y}$ of a Cartesian coordinate system
and the non-orthogonal $\hat{u}$ and $\hat{v}$ used as a convenient
basis in which to solve the Dirac equation in the presence of a
helical potential.}
\end{figure}

Within the continuum theory, we note that this problem can be
reduced to a lattice periodic problem in a twisted coordinate system
that winds with the period of the external potential. We choose our
coordinates so that the $x$-axis lies along the tube and the
$y$-axis wraps around the circumference in a counterclockwise
direction (in this way the $y$ coordinate agrees with the usual
azimuthal angle of radial coordinates).  We let the nanotube radius
be $R$ and the pitch of the helix be $c$, as depicted in
Fig.~\ref{geometry}. The Dirac Hamiltonian in the presence of an
external field becomes
\begin{equation}\label{eq01}
H_K= -i\hbar v_f \vec{\sigma} \cdot \vec{\nabla} + V(x,y),
\end{equation}
where $v_f$ is the Fermi velocity, $\vec{\sigma} = \sigma_x \hat{x}
+ \sigma_y \hat{y}$ are the Pauli matrices, and $V(x,y)$ is the
applied helical potential.  The envelope function is subject to the
quasi-periodic boundary condition $\Psi(x,y+2\pi R) = \exp{(2\pi i
\delta)} \Psi(x,y)$, where $\delta = \pm 1/3$ for semiconducting
tubes and $\delta = 0$ for metallic tubes.\cite{aa1993,mk1997}

The helical symmetry of the external potential implies that
$V(x,y)$ is a function of the single variable $\omega = y/R \pm
2\pi x/c$, where the plus (minus) sign applies to left-handed
(right-handed) helices, and $V(\omega+2\pi) = V(\omega)$. We will
capture the essential physics of the system by investigating the
effects of the lowest Fourier mode of the potential.  We let
$V(x,y) = A \cos{[(y-Px)/R]}$, where $A$ is the potential strength
and $P=\pm 2\pi R/c$ is a convenient dimensionless parameter that
characterizes the geometry of the helical potential, with $P>0$
for right-handed helices and $P<0$ for left-handed helices.

In Appendix \ref{append01}.1 we investigate the effects of higher
Fourier terms and show that for the intravalley scattering
considered below the higher Fourier terms are unimportant.  However,
higher Fourier terms can vary rapidly on the scale of a lattice
constant, and rapidly varying potentials can lead to significant
intervalley scattering.\cite{an1998}  Intervalley scattering
involves a large momentum transfer and in general the Coulomb
potential is unscreened at large momenta.  Such an unscreened
interaction gives higher order Fourier coefficients that are larger
than the lowest screened Fourier coefficient, and we might expect
that intervalley scattering will be at least as important as
intravalley scattering. However, in Appendix \ref{append01}.2 we
show that in general the intervalley scattering matrix elements are
either kinematically forbidden or small compared with the first
screened coefficient.  In the very few cases where intervalley
scattering may be important, we show that the matrix elements are
about the same size as the intravalley elements and do not
appreciably alter any of the results below.

The isotropy of the Dirac equation allows us to choose any two unit
vectors as our basis vectors in the tangent plane of the tube. Here,
it is convenient to take as a basis two unit vectors, $\hat{u}$ and
$\hat{v}$, such that $\hat{u}$ points along an equipotential and
$\hat{v}$ follows the usual azimuthal coordinate. Referring to
Fig.~\ref{geometry}, we see that
\begin{eqnarray}\label{eq02}
\hat{u} &=& \frac{\hat{x}}{\sqrt{1+P^2}} +
\frac{P\hat{y}}{\sqrt{1+P^2}},\nonumber\\
\hat{v} &=& \hat{y}.
\end{eqnarray}
With this choice of basis vectors a general vector in the plane is
written as $\vec{r} = u \hat{u} + v \hat{v}$.  The $(u,v)$
coordinates may be obtained from the Cartesian $(x,y)$ coordinates
by
\begin{eqnarray}\label{eq03}
\left(\begin{array}{c} u \\ v \end{array}\right) =
\left(\begin{array}{cc} \sqrt{1+P^2} & 0 \\ -P & 1 \end{array}
\right) \left(\begin{array}{c} x\\y\end{array} \right).
\end{eqnarray}
In this basis the helical potential is a function of only the $v$
coordinate, $V(v) = A \cos{(v/R)}$, and the Dirac equation becomes
\begin{eqnarray}\label{eq04}
-i \hbar v_f \left(\sqrt{1+P^2} \sigma_x \partial_u + (\sigma_y - P
\sigma_x) \partial_v \right) \Psi \nonumber \\ + A\cos{(v/R)} \Psi =
E \Psi.
\end{eqnarray}

With this choice of spatial basis vectors we are forced to choose
non-orthogonal reciprocal lattice basis vectors.  We choose our
reciprocal basis vectors, $\vec{q}_u$ and $\vec{q}_v$, such that
for a general wavevector, $\vec{k} = k_u \vec{q}_u + k_v
\vec{q}_v$, the dot product with a general spatial vector is given
by $\vec{k} \cdot \vec{r} = k_u u + k_v v$.  This is accomplished
using the reciprocal space (non-unit) vectors $\vec{q}_u =
\sqrt{1+P^2} \hat{x}$ and $\vec{q}_v = -P\hat{x} + \hat{y}$.  With
this basis the $(k_u,k_v)$ coordinates are obtained from the usual
$(k_x,k_y)$ coordinates by the transformation
\begin{eqnarray}\label{eq05}
\left(\begin{array}{c} k_u \\ k_v \end{array}\right) =
\frac{1}{\sqrt{1+P^2}} \left(\begin{array}{cc} 1 & P \\
0 & \sqrt{1+P^2}
\end{array} \right) \left(\begin{array}{c} k_x\\k_y\end{array} \right).
\end{eqnarray}

If we now write $\Psi(u,v) = \exp{(ik_u u)} \psi(v)$, the Dirac
equation becomes
\begin{eqnarray}\label{eq06}
\left(\sqrt{1+P^2}(k_u R)\sigma_x - i R (\sigma_y - P \sigma_x)
\partial_v\right) \psi \nonumber \\  + a \cos{(v/R)} \psi = \epsilon
\psi,
\end{eqnarray}
where $a = A/\Delta_0$ and $\epsilon = E/\Delta_0$ are dimensionless
measures of the potential strength and the energy, respectively, and
$\Delta_0 = \hbar v_f /R$ is a convenient unit of energy.  The above
Hamiltonian has the property that $H_K(-k_u,-P) = \sigma_y
H_K(k_u,P) \sigma_y$, which shows that the spectrum satisfies
$\epsilon(k_u,P) = \epsilon(-k_u,-P)$.  We may therefore only
consider right-handed helices and take $P \geq 0$ in all that
follows.

When $P=0$ Eq.~(\ref{eq06}) also describes a nanotube immersed in
a constant perpendicular electric field, a system studied
previously by Novikov and Levitov.\cite{nl2006,nlarx}  They found
that for semi-conducting tubes the band gap was unaffected by the
applied field until a critical field strength was reached, above
which the band gap closes.  For metallic tubes there exists a
critical field above which the Fermi velocity changes sign and the
Fermi surface fractures.  We will investigate similar effects for
$P \neq 0$ below.

Novikov and Levitov used a chiral gauge transformation to show that,
for the $P=0$ system, the spectrum at $k_x = 0$ is unaffected by the
applied field.  For the $P \neq 0$ system we employ a similar
transformation,
\begin{equation}\label{eq07}
\mathcal{T} = \exp{\left( i a (\sigma_y - P \sigma_x)
\frac{\sin{(v/R)}}{\sqrt{1+P^2}}\right)},
\end{equation}
which converts the Hamiltonian to
\begin{eqnarray}\label{eq08}
H_K^\prime &=& \mathcal{}T H_K \mathcal{T}^{-1} \nonumber \\ &=& - i
R (\sigma_y - P \sigma_x)
\partial_v + \sqrt{1+P^2}(k_u R) \mathcal{T} \sigma_x \mathcal{T}^{-1}. \nonumber \\
\end{eqnarray}
Thus, the spectrum of Eq.~(\ref{eq06}) at $k_u = 0$ is unaffected by
the applied helical potential.

When $k_u = 0$ solutions of Eq.~(\ref{eq06}) are of the form
\begin{equation}\label{eq09}
\psi(v) = f(v)\left(\begin{array}{c} 1 \\ \pm
\frac{P-i}{\sqrt{1+P^2}} \end{array}\right) \exp{\left(\pm i \frac{a
\sin{(v/R)}}{\sqrt{1+P^2}} \right)},
\end{equation}
where $f(v)$ is independent of $a$.  This shows that the applied
field only alters the phase of the wavefunction at $k_u=0$, and we
use this fact to elucidate another protected quantity in the
spectrum of Eq.~(\ref{eq06}).  The $k_u$-space energy gradient is
given by
\begin{equation}\label{eq10}
\frac{\partial \epsilon}{\partial k_u} =\langle \psi |
\frac{\partial H_K}{\partial k_u} |\psi \rangle,
\end{equation}
provided both $\partial \epsilon/\partial k_u$ and $\partial
|\psi\rangle /\partial k_u$ are well defined.  Note that in general
these derivatives are not well defined at degenerate points in the
spectrum.  For non-degenerate $k_u$ points, if we write the
wavefunction as
\begin{equation}\label{eq11}
\psi(v) = \left(\begin{array}{c} \phi(v) \\ \theta(v)
\end{array}\right),
\end{equation}
then Eq.~(\ref{eq10}) becomes
\begin{equation}\label{eq12}
\frac{\partial \epsilon}{\partial k_u} = R \sqrt{1+P^2} \int dv
\left(\phi^* \theta + \theta^* \phi \right).
\end{equation}
By Eq.~(\ref{eq09}), the right hand side is independent of $a$ at
$k_u =0$, which shows that $\partial \epsilon/\partial k_u |_{k_u =
0}$ is unaffected by the external helical potential.  We therefore
find that the longitudinal velocity, given by
\begin{equation}\label{eq12a}
{\rm v} = \frac{1}{\hbar}\frac{\partial E}{\partial k_x} =
\frac{\Delta_0}{\hbar \sqrt{1+P^2}} \frac{\partial
\epsilon}{\partial k_u},
\end{equation}
is unaffected by the helical potential at $k_u = 0$.

\section{\label{sec3} Solutions}

\subsection{\label{subsec3.1} Zero Field Solutions}

\begin{figure}
\includegraphics[width=80 mm]{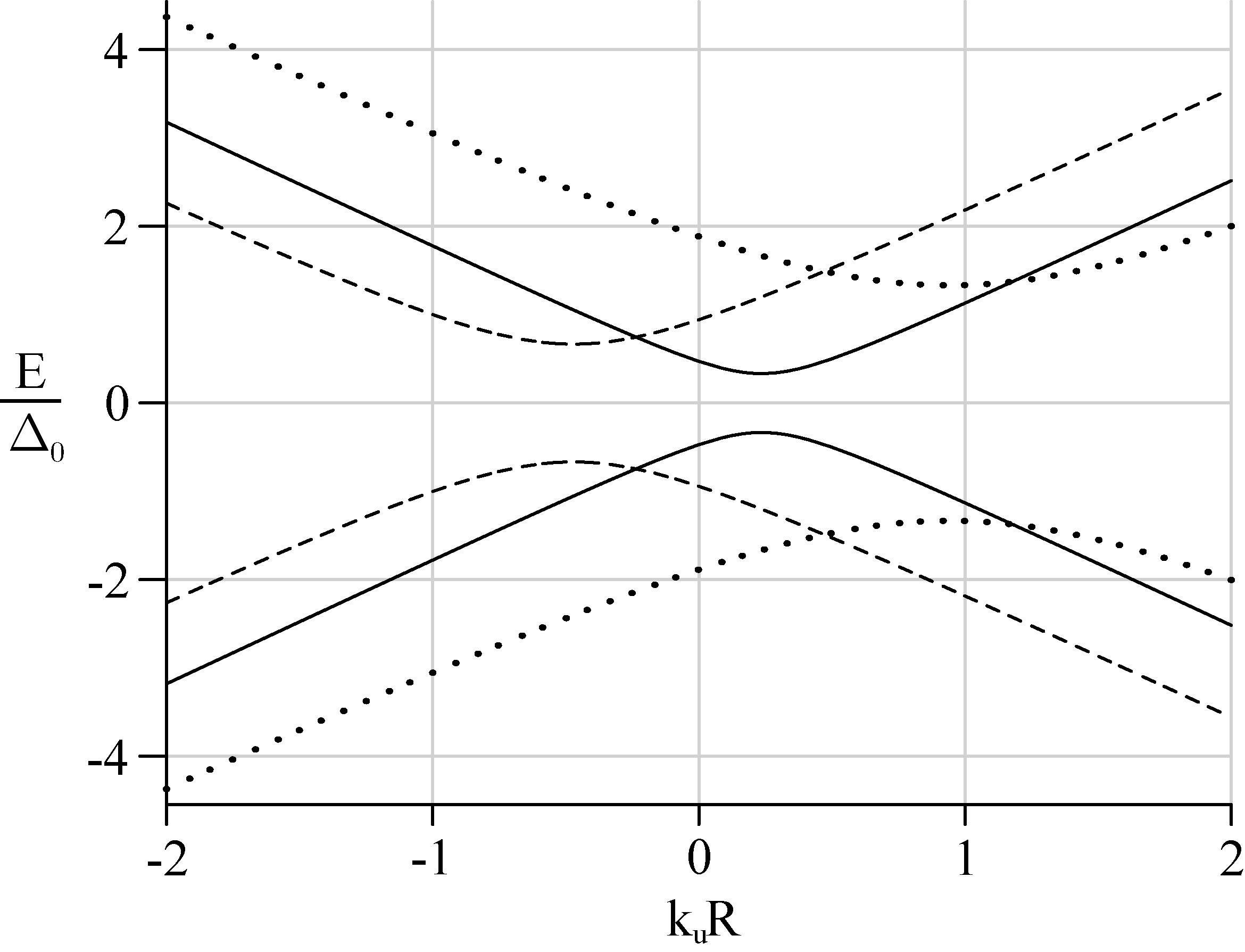}
\caption{\label{freesol}  The three highest valence bands and the
three lowest conduction bands of a semiconducting nanotube ($\delta
= 1/3$) plotted as a function of $k_u R$.  The energy bands are
plotted for a free nanotube ($a=0$) studied in a twisted coordinate
system appropriate for an applied potential with $P=1$. The allowed
energies, like all measurable quantities, are unaffected by the
change in coordinate system.  The apparent differences between this
spectrum and the usual free nanotube spectrum, such as the shift in
the location of band minima, arise because the energies are plotted
as a function of $k_u R$, defined for the twisted coordinate system,
and not the usual $k_x R$.}
\end{figure}

The helical band structure of Eq.~(\ref{eq06}) is  different than
that produced by the ordinary massless Dirac theory on the cylinder
both because of our choice of nonorthogonal coordinates and because
of the applied helical potential. Only the latter change is
physical, and in order to separate the two effects it is necessary
to examine the field free solutions of Eq.~(\ref{eq06}). The
appropriate boundary condition on $\psi(v)$ is obtained by noting
that $\hat{v} = \hat{y}$, so that $\psi(v+2\pi R) = \exp{(2\pi i
\delta)}\psi(v)$.  The field free energies are given by
\begin{equation}\label{eq13}
\epsilon_{\alpha,m}^0 = \alpha \sqrt{\left[\sqrt{1+P^2} (k_u R) - P
(m+\delta)\right]^2 + (m+\delta)^2},
\end{equation}
where the band index $m$ is any integer and $\alpha = \pm 1$ denotes
a conduction or valence band.  The first few energy levels for a
semiconducting CNT are plotted in Fig.~\ref{freesol}.  The band
gaps, like all observable quantities, are unchanged by our choice of
coordinate system.  However, there are two noticeable differences
when the free nanotube spectrum is plotted as a function of $k_u R$,
defined for the twisted coordinated system, compared to the spectrum
when plotted as a function of the usual $k_x R$.  First, the band
minima are shifted to positive (negative) $k_u$ values for $m>0$
($m<0$).  The locations of the new band minima are given by $(k_u
R)_{\rm min} = P (m+\delta)/\sqrt{1+P^2}$.  Secondly, as $k_u R
\rightarrow \infty$, the slope of each band goes as $\sqrt{1+P^2}$ .

An intriguing feature of the band structure of semi-conducting CNTs
is that the bands are shifted so that the magnitude of the slope of
all bands is the same at $k_u =0$, namely $\left| \left.\partial
\epsilon_{\alpha,m}/\partial k_u \right|_{k_u = 0} \right| = P R$.
 As these are the protected locations in the spectrum, it is those
states with $k_u =0$ and with longitudinal velocity ${\rm v} = \pm P
v_f/\sqrt{1+P^2}$ that are unaffected by the applied potential. The
same holds true for the energy bands of metallic tubes except for
the $m=0$ bands. These bands, with energy $\epsilon_{\alpha,0} =
\alpha \sqrt{1 + P^2} (k_u R)$, remain degenerate at $k_u = 0$. From
the analysis at the end of Sec.~\ref{sec2}, the slope of these bands
may be changed by the applied field.  These changes will be
investigated below.

\subsection{\label{subsec3.2} Non-zero Field Solutions}

\subsubsection{\label{subsubsec3.2.1} Semiconducting Nanotubes}

\begin{figure}
\includegraphics[width=80 mm]{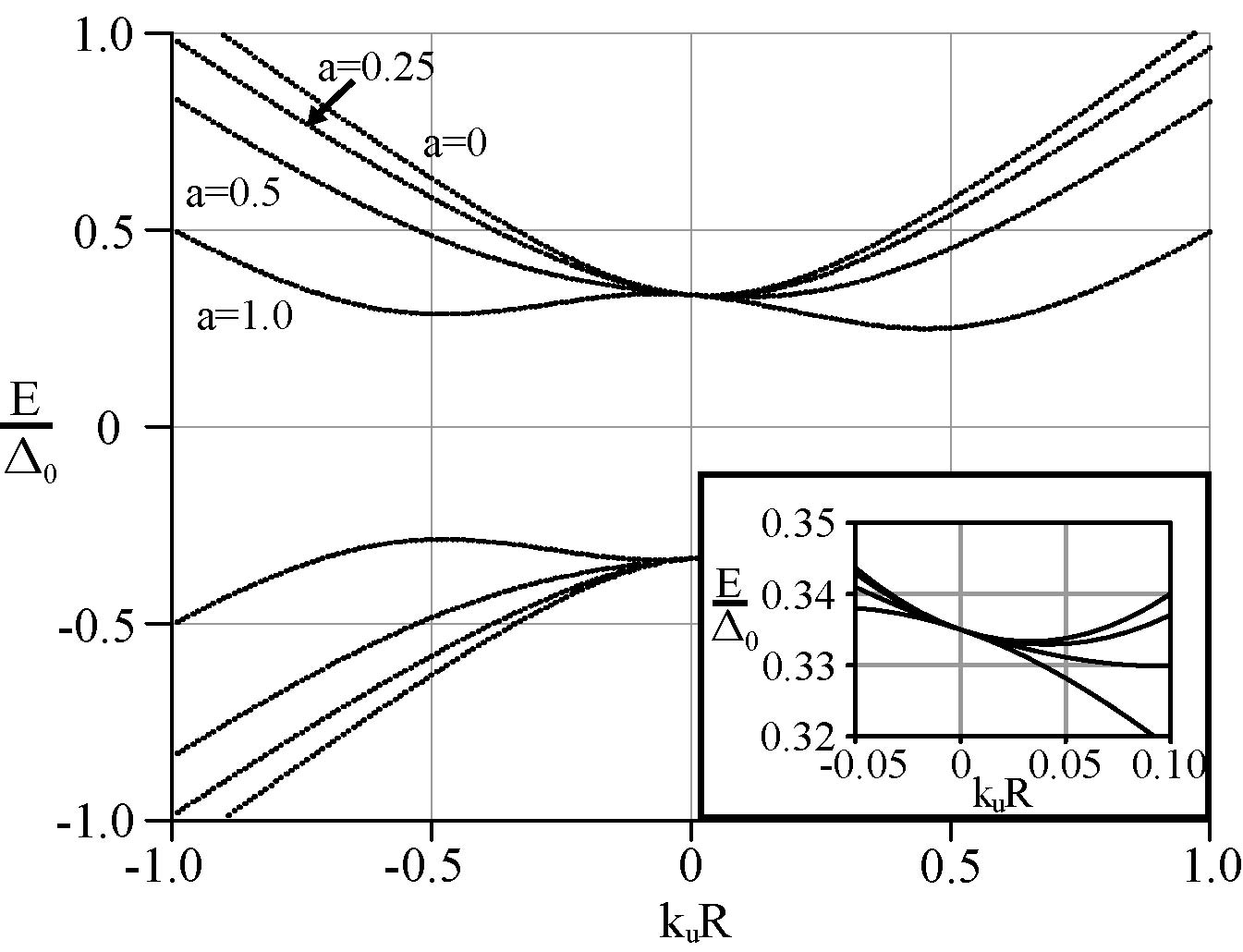}
\caption{\label{firstset}  The lowest conduction band and highest
valence band as function of $k_u R$ for a semi-conducting CNT with
$\delta = 1/3$ and $P = 0.1$, drawn for several values of $a$ as
labeled in the figure.  The inset is an expanded view of the lowest
conduction bands in the area around $k_u R = 0$, and shows that the
band gap closes for every non-zero $a$ considered here.}
\end{figure}

\begin{figure}
\includegraphics[width=80 mm]{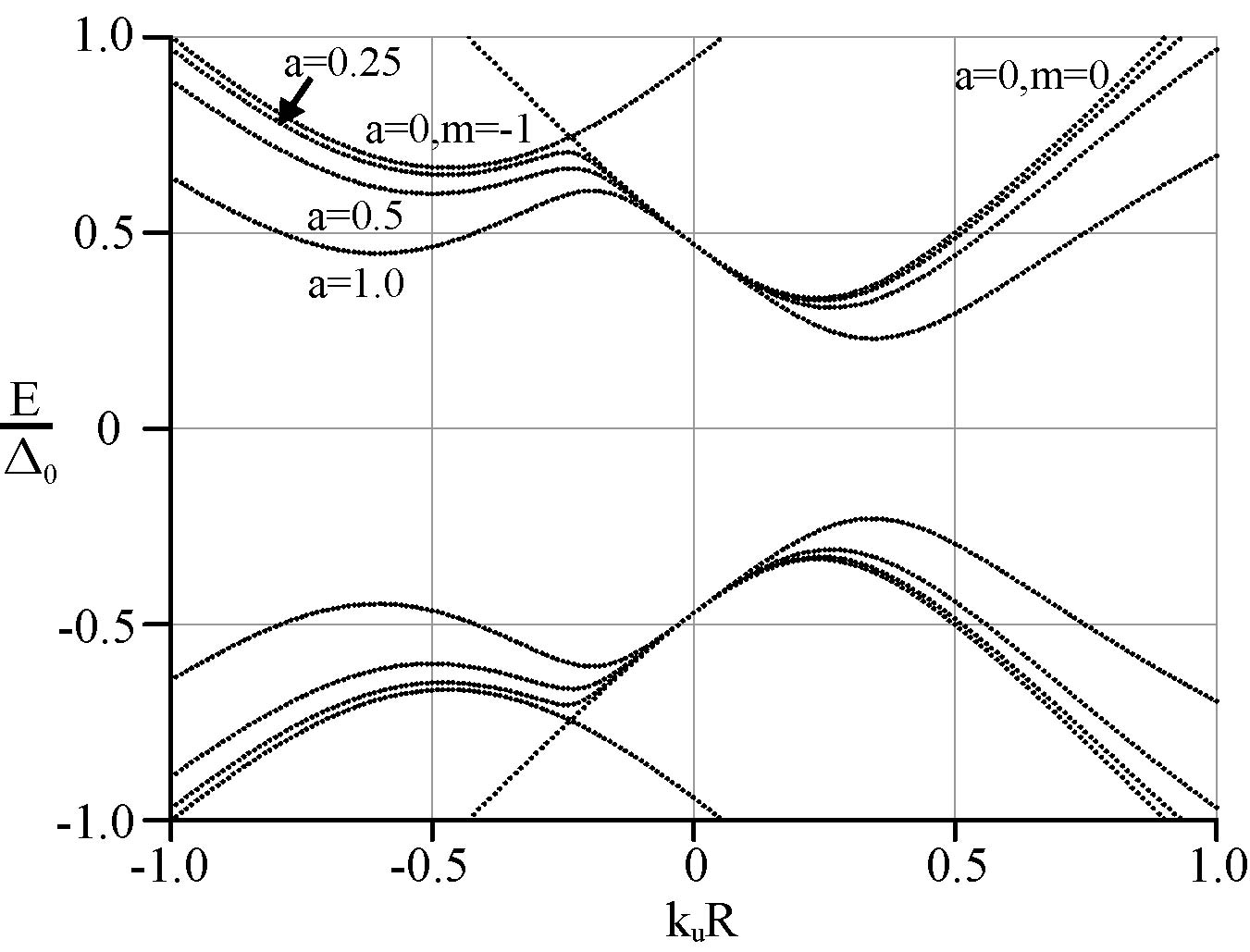}
\caption{\label{secondset}  The lowest conduction band and highest
valence band as function of $k_u R$ for a semi-conducting CNT with
$\delta = 1/3$ and $P = 1.0$, drawn for several values of $a$ as
labeled in the figure.}
\end{figure}

\begin{figure}
\includegraphics[width=80 mm]{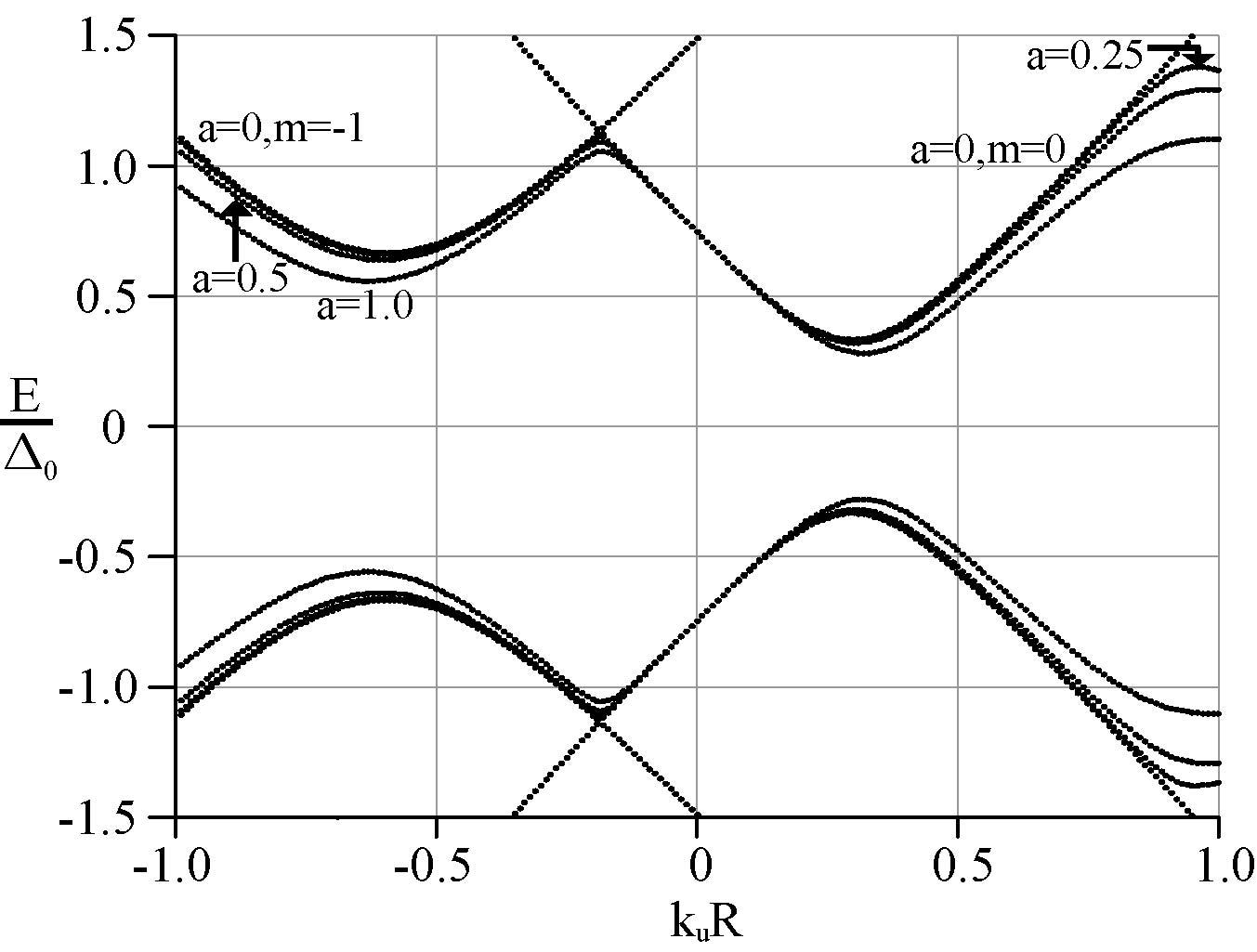}
\caption{\label{thirdset}  The lowest conduction band and highest
valence band as function of $k_u R$ for a semi-conducting CNT with
$\delta = 1/3$ and $P = 2.0$, drawn for several values of $a$ as
labeled in the figure.}
\end{figure}

The spectrum of Eq.~(\ref{eq06}) for semi-conducting nanotubes was
determined numerically for many values of the field strength, $a$,
and the dimensionless geometric parameter, $P$.  A representative
sample of results is shown in Figs.~\ref{firstset} - \ref{thirdset},
where we plot the lowest conduction band and highest valence band
for three different values of $P$ and three different non-zero
values of $a$. The most striking result of these solutions is that,
unlike the $P=0$ case, the band gap closes for any value of $a$.
This is shown in the inset in Fig.~\ref{firstset} and will be
confirmed analytically below. Although there is no critical field to
be applied before the band gap closes, our results suggest that, at
least for $P \lesssim 0.2$, there exists a critical field above
which a second local minimum appears in the band structure. This is
most clearly seen in Fig.~\ref{firstset}, where the $a=1.0$ band
develops a second local minimum at about $k_u R = -0.48$.  For
larger $P$ there is significant mixing of the $m=0$ and $m=-1$ bands
in the region where a second local minimum would develop, and the
existence of a critical field in these cases is less certain.
Nevertheless, for all values of $a$ and $P > 0$ there exists only
one global minimum. The location of the band minimum, $(k_u R)_{\rm
min}$, increases as $a$ increases.

\begin{figure}
\includegraphics[width=80 mm]{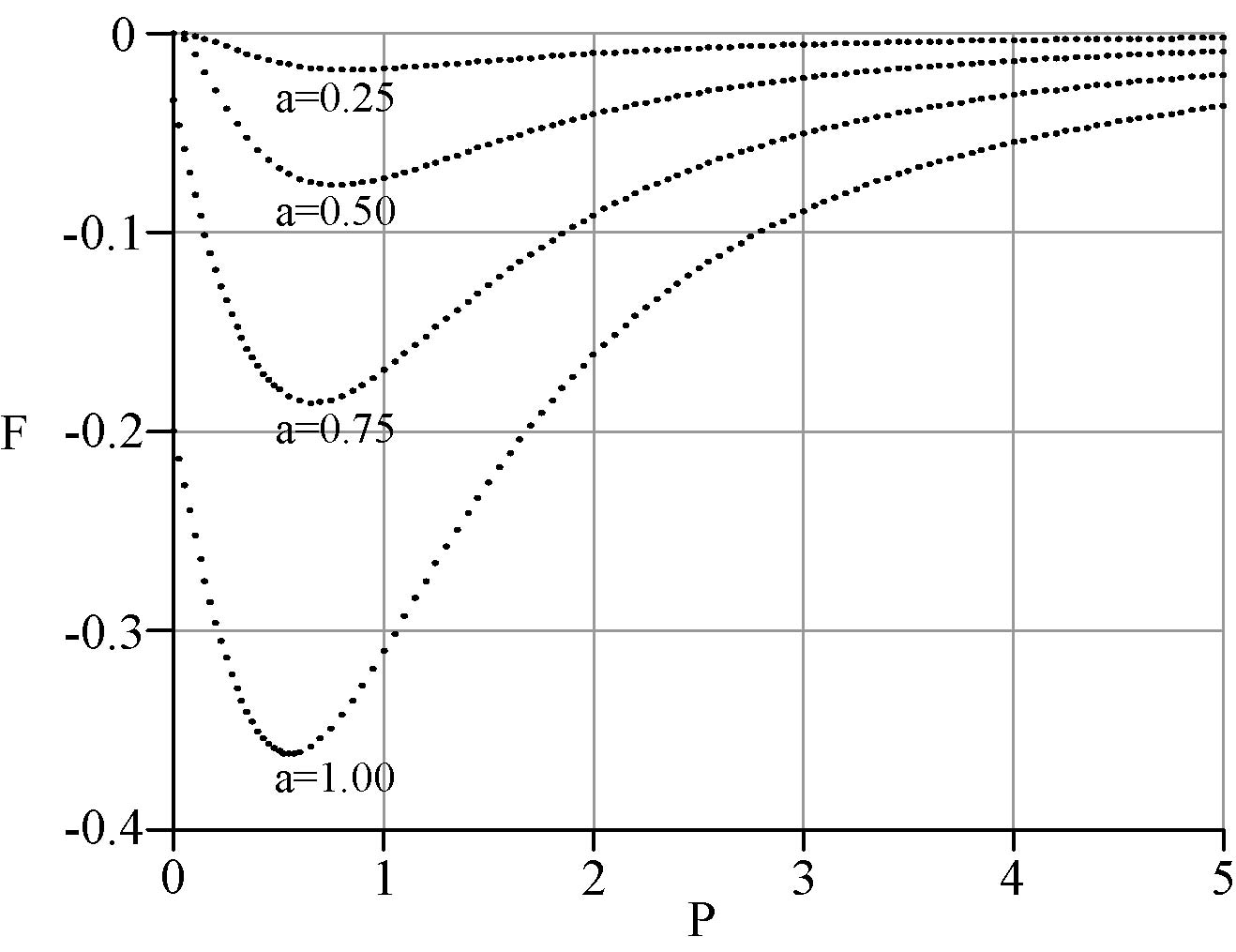}
\caption{\label{bandgap}  The fractional change in the band gap,
$F$, plotted as a function of $P$ for several values of $a$.  The
band gap is unchanged as $P \rightarrow \infty$, as discussed in the
text.  As $P \rightarrow 0$ there are two behaviors:  the band gap
is unchanged if $a < a_c$, but closes for $a > a_c$.  For all $a$
there exists a $P$ on the order of unity that produces the largest
change in the band gap.}
\end{figure}

The fractional change in the band gap is given by $F =
[\Delta(P,a)-\Delta(P,0)]/\Delta(P,0)$, where $\Delta(P,a)$ is the
minimum value of the conduction band for a given $P$ and $a$, and
$\Delta(P,0) = 1/3$ is the minimum value of the unperturbed
conduction band. $F$ is plotted in Fig.~\ref{bandgap} as a function
of $P$ for several values of $a$. For all $a$, $F$ has a maximum for
a $P$ on the order of unity and decays relatively rapidly to zero as
$P \rightarrow \infty$. This may also be seen by comparing
Figs.~\ref{secondset} and \ref{thirdset}, which show that doubling
$P$ significantly decreases the effect of the applied field.  This
behavior is most easily understood by considering the tight-binding
model.  The limit $P \rightarrow \infty$ corresponds to taking the
pitch to zero, at which point the helix collapses into a uniform
cylinder.  The potential is no longer spatially varying on the tube
surface, and its only effect is to provide a uniform background
potential. Such a uniform change may be eliminated by redefining the
zero of energy, and thus cannot affect the band gap.

As $P \rightarrow 0$ the system asymptotically approaches the system
studied by Novikov and Levitov, and their results apply. At $P=0$
the band gap is unaffected by the applied field if $a \leq a_c
\approx 0.6215$, and it closes if $a > a_c$.  These two behaviors
are evident in Fig.~\ref{bandgap}: for $a \leq a_c$, $F$ rapidly
approaches zero as $P\rightarrow 0$, while for $a > a_c$, $F$
approaches an $a$-dependent constant less than zero.

To investigate the size of the band gap for small $a$ we employ
non-degenerate perturbation theory.  The perturbation connects band
$(\alpha,m)$ to bands $(\pm,m+1)$ and $(\pm,m-1)$, and the perturbed
energies are given by
\begin{equation}\label{eq14}
\epsilon_{\alpha,m} = \epsilon_{\alpha,m}^0 +
\left(\frac{a}{2}\right)^2 \frac{G_m}{\epsilon_{\alpha,m}^0},
\end{equation}
where $\epsilon_{\alpha,m}^0$ is given by Eq.~(\ref{eq13}) and
\begin{equation}\label{eq15}
G_m=\frac{4(k_u R)^2}{4\left[P(k_u R) - \sqrt{1 + P^2}
(m+\delta)\right]^2 - (1+P^2)}.
\end{equation}
Because of the many degeneracies in the spectrum at high energy(see
Fig.~\ref{freesol}), the effects of the chiral potential on the
entire spectrum cannot be studied within a non-degenerate
perturbation theory. However, there is never a degeneracy at the
{\it band minimum} of the $m=0$ band, and Eq.~(\ref{eq14})
accurately reproduces the band gaps found in numerical results. The
fractional change in the band gap may be estimated by evaluating
Eq.~(\ref{eq14}) at the unperturbed band minimum, which gives
\begin{equation}\label{eq16}
F_{\alpha,0} = \alpha \left(\frac{a}{2}\right)^2 \frac{4P^2
|\delta|}{4\delta^2 - (1+P^2)^2}.
\end{equation}
Notice that as $(1+P^2) > 2|\delta|$ for all $P$, the band gap
always closes.  The actual band gap is in fact slightly smaller than
predicted by Eq.~(\ref{eq16}) because the unperturbed band minimum
is not the true band minimum.  Nevertheless, the conclusion that the
band gap closes for any $a>0$ remains valid.

To investigate the existence of a critical field and the development
of a second local minimum we employ non-degenerate perturbation
theory in small $k_u R$.  The perturbation connects conduction bands
to valence bands, and the perturbed energies are given by
\begin{eqnarray}\label{eq17}
\epsilon_{\alpha,m} &=& \alpha |m+\delta| \sqrt{1+P^2} - \alpha
\sgn( m+\delta) P (k_u R) \nonumber \\ & &+ \alpha \sgn(m+\delta)
\frac{(k_u R)^2 H_m}{\sqrt{1+P^2}},
\end{eqnarray}
where $\sgn(x)$ is the sign function and
\begin{equation}\label{eq18}
H_m = \sum_{n=-\infty}^\infty J_n^2\left(\frac{2
a}{\sqrt{1+P^2}}\right) \frac{2 (m+\delta)}{4(m+\delta)^2 - n^2}.
\end{equation}
For $\delta = 1/3$, $H_0$ switches sign from positive to negative at
$a_c \approx 0.6215 \sqrt{1+P^2}$.  For $a>a_c$ the curvature at
$k_u = 0$ is negative and a second local minimum develops to the
left of the origin, as deduced from the numerical results.

\subsubsection{\label{subsubsec3.2.2} Metallic Tubes}

\begin{figure}
\includegraphics[width=80 mm]{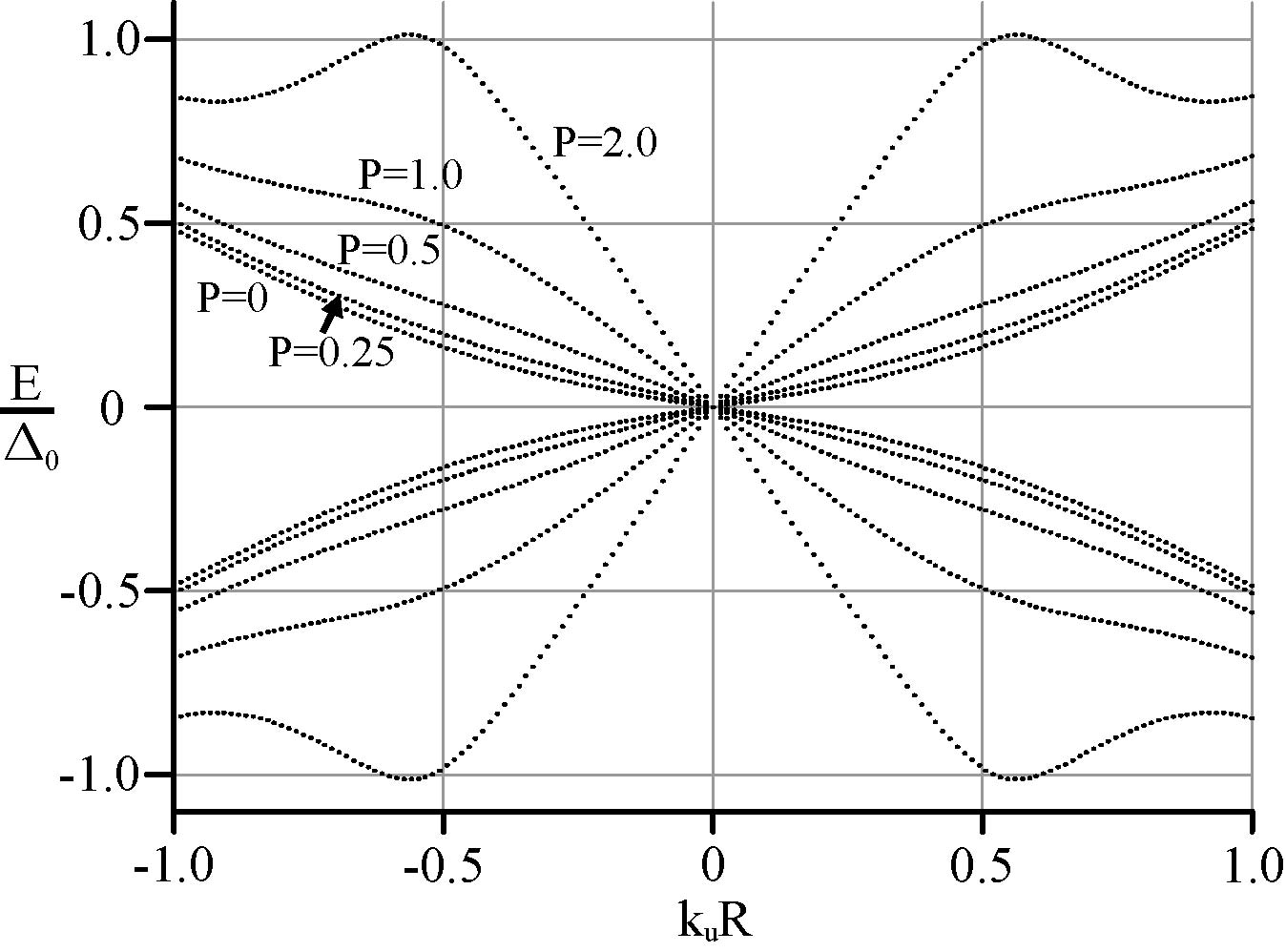}
\caption{\label{metal1}  The low energy bands of a metallic nanotube
as a function of $k_u R$ for $a = 1.0 < a_{cm}$ for several values
of $P$ as labeled in the figure.  The primary effect of the field is
to modify the Fermi velocity.}
\end{figure}

\begin{figure}
\includegraphics[width=80 mm]{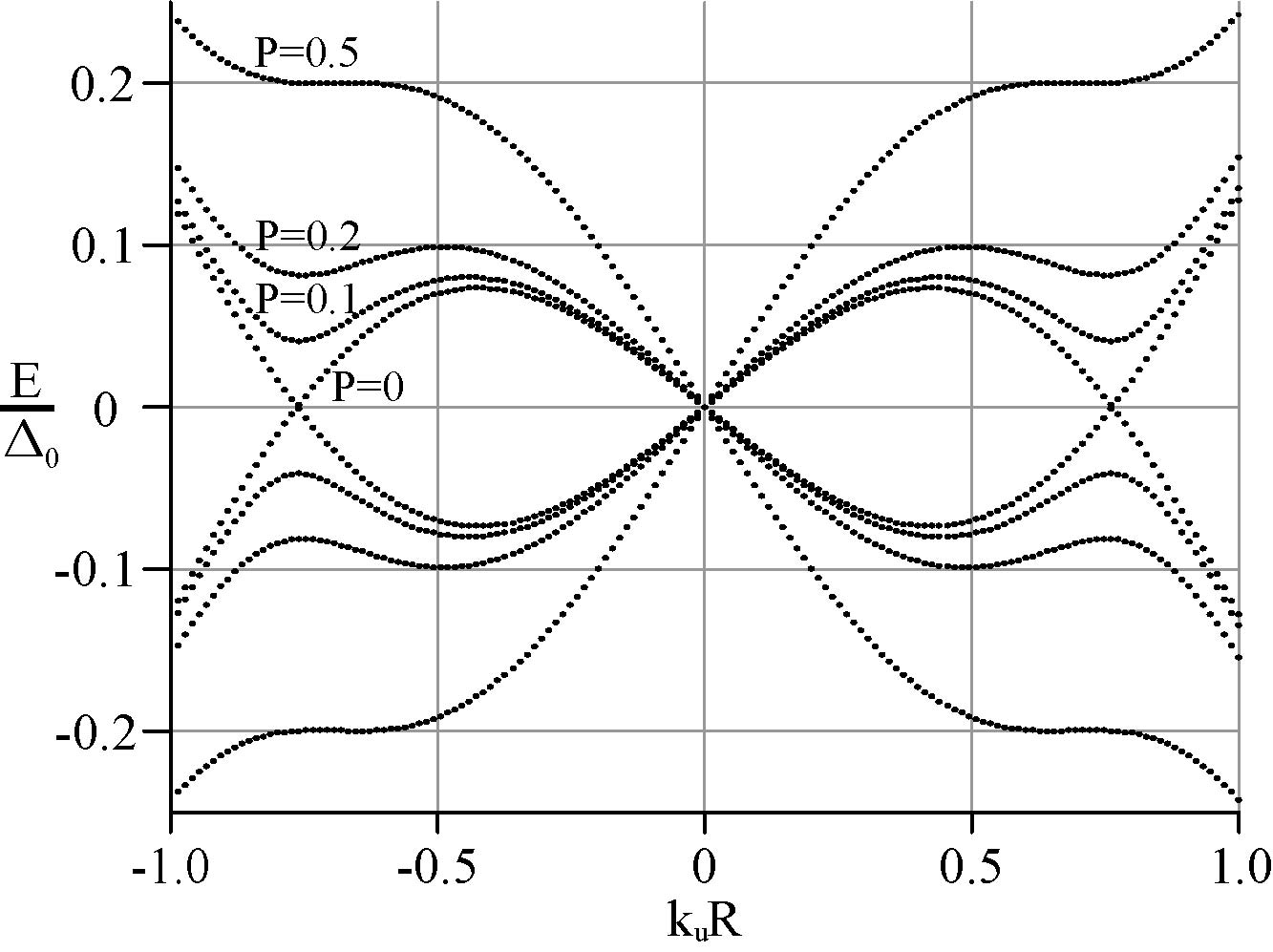}
\caption{\label{metal2}  The low energy bands of a metallic nanotube
as a function of $k_u R$ for $a = 1.5 > a_{cm}$ for several values
of $P$ as labeled in the figure.  For $P = 0$ the Fermi surface
fractures, for $P > 0$ the Fermi surface is unchanged.}
\end{figure}

The low energy spectrum of Eq.~(\ref{eq06}) for metallic tubes is
shown in Figs.~\ref{metal1} and \ref{metal2} for two different
values of $a$ and several values of $P$.  As discovered by Novikov
and Levitov, for $P=0$, as the field is turned on the Fermi velocity
decreases and the band acquires a non-zero curvature away from $k_u
= 0$.  The Fermi velocity goes to zero at a critical field strength,
$a_{cm} \approx 1.203$.  When $a > a_{cm}$ the Fermi velocity
switches sign and the Fermi surface fractures, as shown in
Fig.~\ref{metal2}.

The spectrum for fixed $a < a_{cm}$ is plotted in Fig.~\ref{metal1}
for several $P$ values.  As $P$ increases the slope of the energy
bands at $k_u = 0$ increases. However, note that as $v_f =
\partial \epsilon/\partial k_x = (1/\sqrt{1+P^2})\partial
\epsilon/\partial k_u$, the increase in the slope does not
necessarily translate into an increase in the Fermi velocity.  In
fact, for very large $P$ the slope increases linearly in $P$, which
implies an unchanged Fermi velocity in the limit $P \rightarrow
\infty$. This effect could have been anticipated from the analysis
of the semiconducting band gap, where we argued that the effect of
the chiral potential must disappear in the limit $P \rightarrow
\infty$. As $P$ increases the higher energy states mix with the $m =
\pm 1$ bands and two symmetric local minima develop in the lowest
band.

The spectrum for fixed $a > a_{cm}$ is plotted in Fig.~\ref{metal2},
where it is seen that the effect of a non-zero $P$ is to open up
gaps where the $P=0$ bands crossed the Fermi surface at non-zero
$k_u$. For any $P>0$ the Fermi surface does not fracture, and for
this property $P=0$ is a singular limit.  As $P$ increases the slope
at $k_u = 0$ increases and the local minima at the induced band gaps
at first flatten, then reappear as the lowest band begins to mix
with the $m=\pm 1$ bands. As $P\rightarrow \infty$ the effects of
the chiral potential disappear and the $a>a_{cm}$ system maps
directly onto the $a < a_{cm}$ system.

We study the change in the Fermi velocity using degenerate
perturbation theory on the two states at $k_u = 0$.  The perturbed
energies are given by
\begin{equation}\label{eq19}
\epsilon_{\alpha,0} = \alpha \sqrt{P^2 + J_0^2\left(2
a/\sqrt{1+P^2}\right)} (k_u R).
\end{equation}
For $P=0$ the Fermi velocity vanishes when $2 a = \mu_1 \approx
2.405$, the first zero of the Bessel function.  When $P>0$ the Fermi
velocity is strictly non-zero.  The slope of the bands is an
oscillatory function of $a$, but as $J_0(x) \leq 1$, this
oscillatory behavior will be small for large $P$.  As $P \rightarrow
\infty$ the slope goes as $P$, in agreement with the numerical
results above.

\section{\label{sec4} The Field Strength}

\begin{figure}
\includegraphics[width=80 mm]{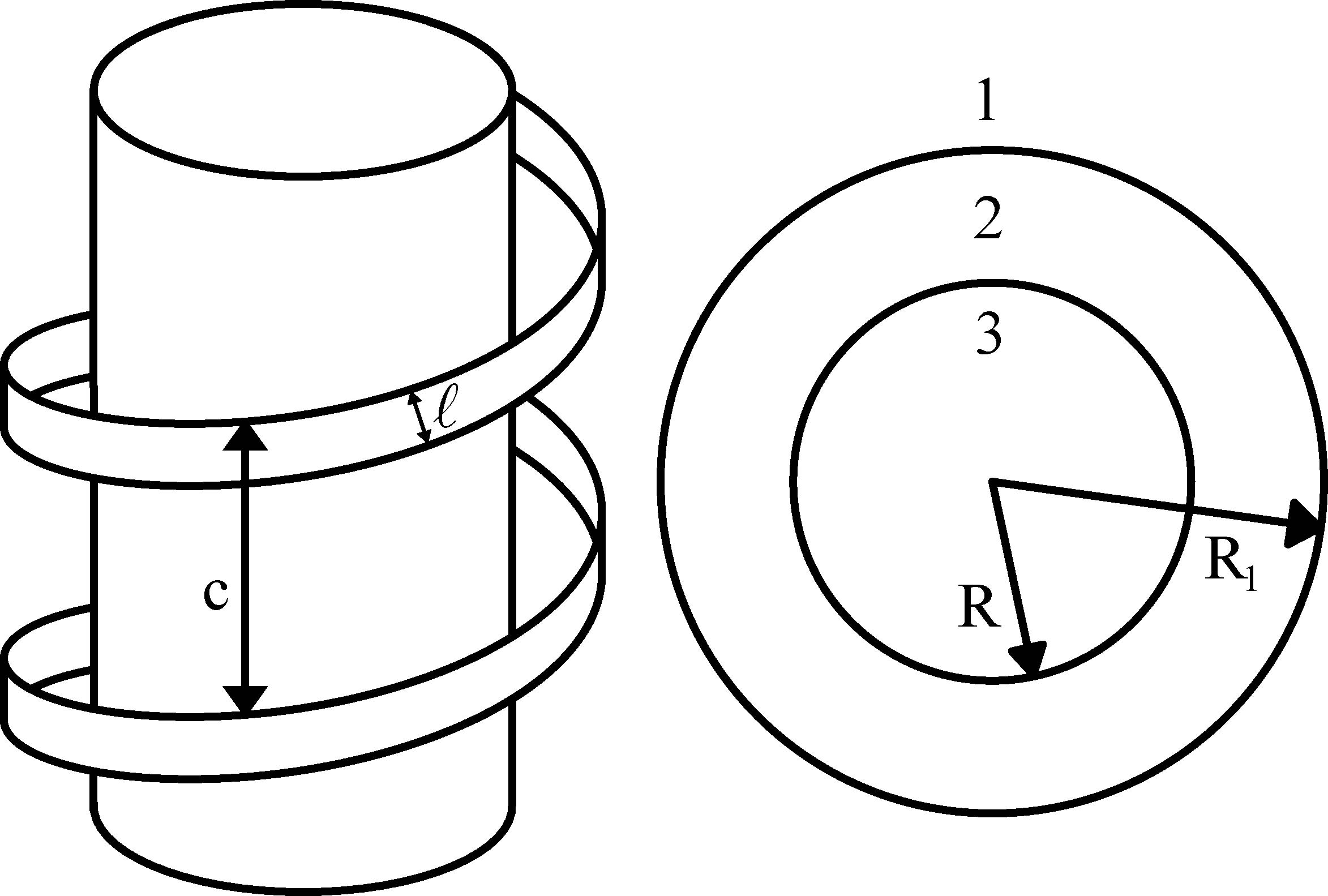}
\caption{\label{ribbon}  On the left is a depiction of our model, a
tube of radius $R$ surrounded by a helical ribbon of radius $R_1$,
width $l$, and surface charge density $\sigma_0$.  On the right is a
cross-sectional view of the system.  The numbers refer to the three
regions of space with different dielectric constants, as discussed
in the text.}
\end{figure}

To a first approximation, the DNA and polymers used to wrap
nanotubes may be considered infinitely long, helically wrapped line
charges.  In order to estimate the magnitude of the field strength
we model the wrapping molecule as a helical ribbon of width $l$,
radius $R_1$, and pitch $c$, with surface charge density $\sigma_0$
(see Fig.~\ref{ribbon}). The field strength is obtained by
evaluating the leading Fourier coefficient of this charge
configuration on the nanotube surface, $r=R$.  At the end of this
calculation we recover the line charge model by taking the limit
$l\rightarrow 0$ with the linear charge density, $\lambda = l
\sigma_0$, held fixed.  In this section we work in SI units.

In a long wavelength theory of the applied fields,  we break space
into three macroscopic regions characterized by different dielectric
constants, as shown in Fig.~\ref{ribbon}. Region $1$, $r>R_1$, is
composed of material outside the helical line charge, the aqueous
solution or other solvent, or possibly vacuum. Region $2$, $R_1 > r
> R$, is composed of the material between the line charge and the
nanotube, including DNA bases or uncharged regions of a wrapping
polymer. Region $3$, $R>r>0$, is composed of the nanotube itself.
Denote the dielectric constant in the $i$th region by $\epsilon_i$.

The helical symmetry of the charge distribution implies that the
potential can be expanded in cylindrical coordinates as
\begin{eqnarray}\label{eq20}
V_i(r,\phi,z) &=& B_0^{(i)} \ln{(r/r_0^{(i)})} + \sum_{n=1}^\infty
\left(A_n^{(i)} I_n(2\pi n r/c) \right. \nonumber \\ &+& \left.
B_n^{(i)} K_n(2\pi n r/c)\right) \cos[n(\phi-2\pi z/c)], \nonumber
\\
\end{eqnarray}
where $i=1,2,3$ denotes the three regions of space and
$A_n^{(i)}$,$B_n^{(i)}$, and $r_0^{(i)}$ are expansion coefficients
to be determined.  The coefficients are determined using the usual
electrostatic boundary conditions,\cite{griffem}
\begin{eqnarray}\label{eq21}
V_> (r=r_i) &=& V_<(r=r_i), \nonumber \\
\left.\epsilon_> \frac{\partial V_>}{\partial r}\right|_{r=r_i}
\!\!\!- \left.\epsilon_< \frac{\partial V_<}{\partial
r}\right|_{r=r_i}\!\!\! &=& -\sigma(r=r_i)/\epsilon_0,
\end{eqnarray}
where the boundary is located at $r=r_i$, $V_>$ and $\epsilon_>$
($V_<$ and $\epsilon_<$) denote the potential and dielectric
constant for $r \geq r_i$ ($r \leq r_i$), and $\sigma$ is the charge
density at the boundary. The charge density at $r=R_1$ is given by
\begin{equation}\label{eq22}
\sigma(\phi,z) = \frac{x_0 \sigma_0}{c} + \sum_{n=1}^\infty \frac{2
\sigma_0}{n \pi} \sin(n\pi x_0/c) \cos[n(\phi - 2\pi z/c)],
\end{equation}
where $x_0 = l \sqrt{1+P_1^2}/P_1$ and $P_1 = 2\pi R_1/c$.

The field strength is obtained from the Fourier expansion by $a = -e
A_1^{(1)} I_1(P)/\Delta_0$, where the electron charge is $-e$. After
taking the limit as the ribbon goes to a line charge, we find
\begin{equation}\label{eq23}
a = \frac{e \lambda}{\pi \hbar v_f \epsilon_3 \epsilon_0} W(R_1,R) =
C W(R_1,R),
\end{equation}
where $C$ is independent of $R$ and $R_1$, and
\begin{equation}\label{eq24}
W = \left(\frac{\epsilon_3}{\epsilon_1}\right)\frac{R \hspace{1 mm}
U(P_1,P)}{K_1^\prime(P_1) S(P_1,P) - (\epsilon_2/\epsilon_1)
K_1(P_1) T(P_1,P)},
\end{equation}
with
\begin{eqnarray}\label{eq25}
U(x,y) &=& \frac{\sqrt{1+x^2}}{x}\frac{K_1(x)
I_1(y)}{1-(\epsilon_3/\epsilon_2)} \left(\frac{K_1(y)}{I_1(y)} -
\frac{K_1^\prime(y)}{I_1^\prime(y)}\right), \nonumber \\
S(x,y) &=& K_1(x) + \frac{I_1(x)}{1-(\epsilon_3/\epsilon_2)}
\left(\frac{\epsilon_3}{\epsilon_2} \frac{K_1(y)}{I_1(y)} -
\frac{K_1^\prime(y)}{I_1^\prime(y)}\right), \nonumber \\
T(x,y) &=& K_1^\prime(x) +
\frac{I_1^\prime(x)}{1-(\epsilon_3/\epsilon_2)}
\left(\frac{\epsilon_3}{\epsilon_2} \frac{K_1(y)}{I_1(y)} -
\frac{K_1^\prime(y)}{I_1^\prime(y)}\right). \nonumber \\
\end{eqnarray}

To determine $C$ we consider specifically the case of an ssDNA-CNT
complex.  The linear charge density of ssDNA is obtained by assuming
each phosphate group on the backbone carries a charge of $-e$, which
gives $\lambda \approx -1.5 e/{\rm nm}$.  The dielectric constant of
a CNT is obtained by using a result from
Refs.~(\onlinecite{blc1995}) and (\onlinecite{nl2006}), that when
immersed in a perpendicular electric field the ratio of the field
strength inside the tube to the applied field is given by $E/E_0 =
1/5$, independent of $R$.  If the nanotube is modeled as a uniform
solid cylinder with dielectric constant $\epsilon_3$ then $E/E_0 =
2/(\epsilon_3 + 1)$, which gives $\epsilon_3 = 9$. Using these
results and $v_f \approx 8 \times 10^5 \hspace{1 mm} {\rm m/s}$, we
find $C \approx -1.8 \hspace{1 mm}{\rm nm}^{-1}$.

The distance between the ssDNA backbone and CNT surface is
independent of nanotube radius, and is given by $R_1 - R \approx
0.6$ nm.  The dielectric constant for region $2$ is difficult to
approximate as it is composed of random DNA bases, solvent molecules
and other dissolved species.  However, provided the pitch is large
compared to the nanotube radius the dominant screening effects
should come from regions $1$ and $3$.  We therefore ignore screening
in region $2$ and set $\epsilon_2 = 1$.

\begin{figure}
\includegraphics[width=80 mm]{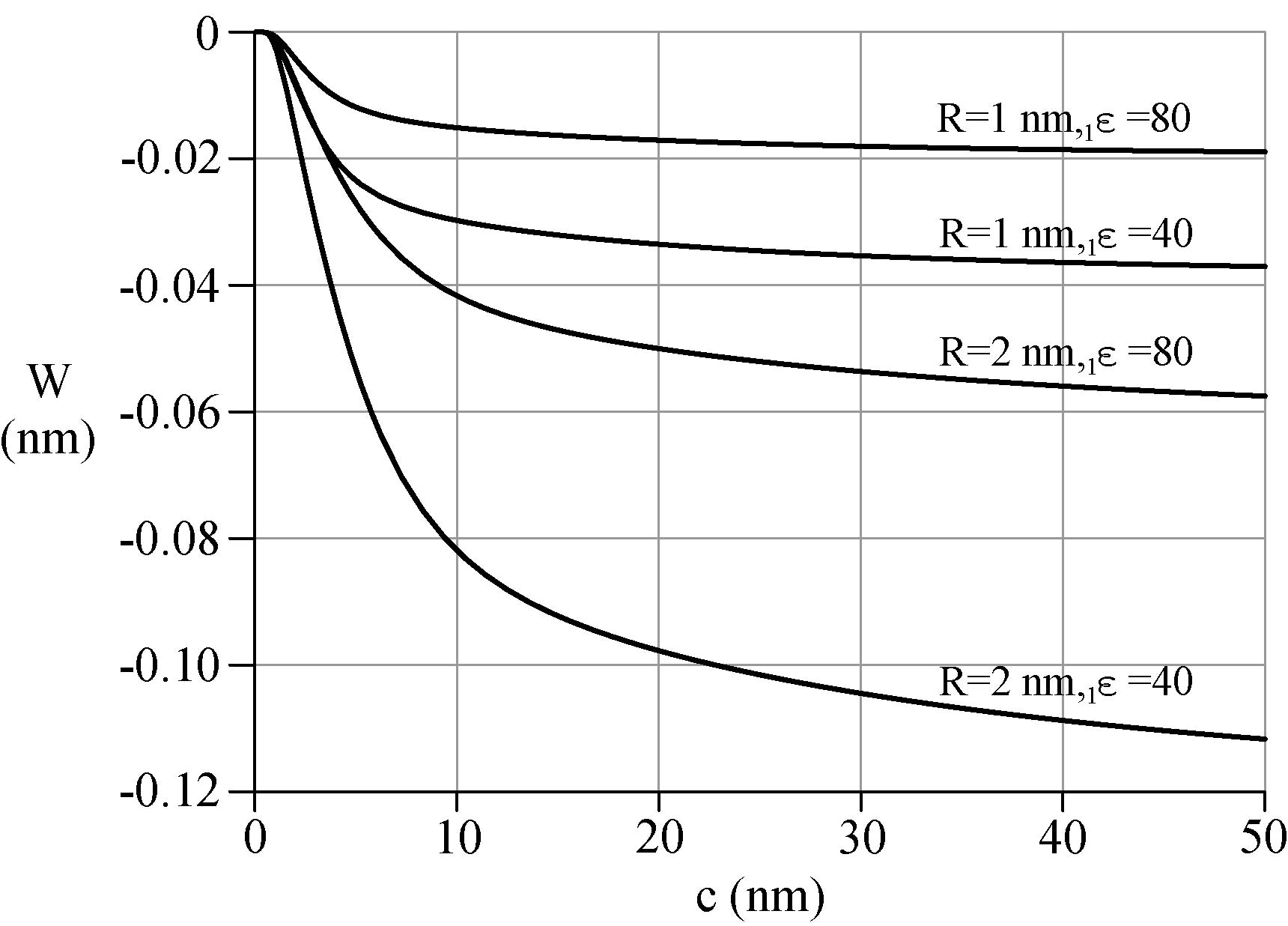}
\caption{\label{wfunction}  The function $W(R,c,\epsilon_1)$ as a
function of $c$ for two values of $R$ and two values of
$\epsilon_1$.}
\end{figure}

With these approximations $W = W(R,c,\epsilon_1)$.  The asymptotic
behavior of $W$ for $R \gtrsim 1$ nm and $\epsilon_1 \gtrsim 5$ is
$W \sim w_0 R/\epsilon_1$, where $w_0$ is a function of $c$.  The
behavior of $W$ as a function of $c$ is shown in
Fig.~\ref{wfunction} for two values of $R$ and $\epsilon_1$.  $W$
goes to zero as $c \rightarrow 0$, and goes asymptotically to a
constant value as $c \rightarrow \infty$. Physically realizable
values of $c$ are in the $1 - 60$ nm range,\cite{gsb2006} and in
this region $W$ is monotonically decreasing but its behavior is not
given by a simple scaling relation.

To obtain an estimate of the field strength we use a CNT with $R =
0.5$ nm and a solution with $\epsilon_1 = 80$, which is
approximately the dielectric constant of water.  In Table~\ref{tab1}
we calculate the field strength and the corresponding fractional
change in the band gap for several values of the pitch.  As $c$
increases $a$ asymptotically approaches $1.1 \times 10^{-2}$, while
$P$ monotonically decreases to zero.  From Fig.~\ref{bandgap} we
expect the magnitude of $F$ will be peaked for a $P$ on the order of
unity and decay to zero for very small and very large pitches.  This
behavior is confirmed in Table~\ref{tab1}, where $c = 5$ nm produces
the largest change in the band gap.  Notice that as the pitch
increases the field strength continues to increase, but the
corresponding decrease in $P$ reduces the effect of the applied
potential.

For the largest fractional change found here, the difference between
the original and perturbed band gap is about $0.01$ meV. We do not
expect a change of this scale to be readily measurable in transport
or optical experiments. There are several possible ways one might
consider to increase the size of this effect.  The first is to use
large radius nanotubes, which will increase the magnitude of the
applied chiral potential. This approach has two problems.
Fundamentally, as $R$ is increased $P$ is also increased, and the
size of the effect decreases with increasing $P$. Practically,
single-walled nanotubes are generally no larger than $1.0$ nm in
diameter, with larger radius tubes unstable to
collapse.\cite{gcg1998}  The second method is to tune $c$ so that
for a given radius nanotube, $P$ is slightly less than unity where
the effect of the helical potential is largest.  For shorter DNA
strands a systematic study of pitch with varying bases is
possible,\cite{zjss2003} but for a long DNA strand with hundreds or
thousands of bases such a study is impractical.  A third method is
to change the wrapping polymer to one with a larger linear charge
density or to one that rests closer to the nanotube surface. The
linear charge density of DNA is already rather large and simply
doubling or tripling the linear charge density will not increase the
size of the effect enough for easy observation. The field strength
increases exponentially as the charge approaches the nanotube
surface, but the decay length is large relative to typical molecular
distances.
 A fourth method is to change the
environment of the DNA-CNT complex. The dielectric constant of water
is unusually large and essentially reduces the field strength by a
factor of $80$. Most organic solvents have dielectric constants in
the range $1 - 5$, which would increase the field strengths
calculated above by factors of $16 - 80$.  The DNA-CNT complex is
remarkably stable, and it may be possible to create the complex in
aqueous solution and then transfer it to an organic solvent.  Other
wrapping molecules have already been used to dissolve CNTs in
organic solvents,\cite{cga2001,sss2001} and one of these might
produce field strengths strong enough to produce a measurable
effect.

\begin{table}
\caption{\label{tab1} Calculated values of $P$, the field strength,
$a$, and the fractional change in the band gap, $F$, for various
values of the pitch, $c$.  The other parameters of the system are
described in the text.}
\begin{ruledtabular}
\begin{tabular}{cccc}
$c$ (nm) & $P$ & $a$ & $F$ \\
\hline 1& 3.14 & $6.6 \times 10^{-4}$  & $-3.6 \times 10^{-8}$\\
2&1.57&$3.7\times 10^{-3}$ & $-2.9 \times 10^{-6}$\\
5&0.63&$7.6\times 10^{-3}$ & $-1.5 \times 10^{-5}$\\
10&0.31&$8.8\times 10^{-3}$ & $-1.0 \times 10^{-5}$\\
25&0.13&$1.0 \times 10^{-2}$ & $-2.7 \times 10^{-6}$\\
50&0.063&$1.0\times 10^{-2}$ & $-7.7 \times 10^{-7}$\\
\end{tabular}
\end{ruledtabular}
\end{table}

\section{\label{sec5} Conclusion}

We have investigated the effect of a helical potential on
semi-conducting and metallic nanotubes.  For semi-conducting
nanotubes the band gap closes for any non-zero field strength.  The
size of the effect is determined by both the field strength, $a$,
and a dimensionless geometrical factor $P$, which is the ratio of
the circumference of the nanotube to the pitch of the helix.   For
each $a$ there exists an optimal $P$ that produces the biggest
change in the band gap. For metallic tubes, the helical potential
decreases the Fermi velocity but does not fracture the Fermi
surface.  Under typical conditions the effect of the helical
potential is probably unobservable, but we understand the scaling
relationship between the size of the band gap and every control
parameter, so in theory it is possible to design a system where
these effects would be detectable in an optical experiment. In some
ways the small size of the effect is encouraging, as it means that
measurements on helically wrapped CNTs give results that are nearly
identical to those of pristine CNTs. Also, a helically wrapped CNT
can be substituted for a pristine CNT in almost any application,
which may make device construction easier.

The theory developed here ignores higher order corrections, such as
curvature effects and higher Fourier terms in the potential.  Simply
expanding the tight-binding graphene Hamiltonian to the next lowest
gradient order generates terms that break the chiral gauge symmetry
and introduce a correction to the spectrum at $k_u =0$.  Such
corrections will modify the functional form of our analytic
expressions; for example, curvature effects will introduce a chiral
angle dependent band gap.  However, curvature effects and higher
order expansion terms introduce corrections that are small compared
to the energies of interest.  The theory developed here also ignores
exciton effects, which are known to be large in CNTs.  Exciton
effects in optical experiments may alter the numerical values
obtained here by $20 - 30 \%$, but they should not significantly
change our general results or alter the conclusions outlined in the
previous paragraph.

This work was supported by the Department of Energy under grant
DE-FG-84ER45118.

\appendix
\section{\label{append01} Higher Fourier Coefficients}

A general chiral potential on the nanotube may be written as
\begin{equation}\label{eqa1}
V(\vec{r})=\sum_{n=1}^\infty V_n \cos(\vec{Q}_n \cdot \vec{r}),
\end{equation}
where $\vec{Q}_n = \pm 2\pi n/c \hspace{1 mm} \hat{x} + n/R
\hspace{1 mm} \hat{y}$.  In our calculations above we only retained
the first Fourier term in order to capture the essential physics of
the system.  In this Appendix we investigate the effect of including
the higher Fourier terms in the Hamiltonian.  Our perturbation
calculations above only included contributions from intravalley
scattering matrix elements, specifically, scattering from a state
near the $\mathbf{K}$ point to another state near the $\mathbf{K}$
point. In section \ref{append01}.1 we show that in all cases higher
Fourier coefficients add a negligible correction to the intravalley
matrix elements.

The higher Fourier terms vary rapidly on the scale of a lattice
constant and contribute to scattering with a large momentum
transfer.  Such scattering may connect states at $\mathbf{K}$ with
those at $\mathbf{K}^\prime$, so we must also consider intervalley
scattering matrix elements in our perturbation expansion.  At large
momentum transfer the Coulomb interaction is unscreened and the
Fourier coefficients in Eq.~(\ref{eqa1}) will be significantly
larger than the corresponding screened coefficients.  Nevertheless,
we show in section \ref{append01}.2 that in almost all cases the
intervalley scattering matrix elements can be neglected compared to
the intravalley scattering matrix elements.  In a few systems with
small radius nanotubes, the intervalley matrix elements may be
nearly as large as the intravalley scattering matrix elements.  We
discuss these few cases below.

\subsection{\label{append01.1} Intravalley Scattering}

\begin{table}
\caption{\label{tab2} Calculated values of higher field strength
Fourier coefficients for the nanotube system studied in
Sec.~\ref{sec4}.}
\begin{ruledtabular}
\begin{tabular}{ccccc}
$c$ (nm) & $a_1$ & $a_2/a_1$ & $a_3/a_1$ & $a_4/a_1$ \\
\hline
2  & 0.0037 & 0.063 & $5.4 \times 10^{-3}$ & $5.2 \times 10^{-4}$ \\
5  & 0.0075 & 0.16  & $3.7 \times 10^{-2}$ & $9.5 \times 10^{-3}$ \\
10 & 0.0088 & 0.20  & $5.6 \times 10^{-2}$ & $1.8 \times 10^{-2}$ \\
50 & 0.0104 & 0.20  & $5.9 \times 10^{-2}$ & $2.0 \times 10^{-2}$ \\
\end{tabular}
\end{ruledtabular}
\end{table}

The dimensionless potential energy appearing in the Hamiltonian is
given by
\begin{equation}\label{eqa2}
U(\vec{r}) = \sum_{n=1}^\infty a_n \cos(\vec{Q}_n \cdot \vec{r}),
\end{equation}
where $a_n = -e V_n/\Delta_0$.  The coefficient $a_1$ was evaluated
in Sec. \ref{sec4}.  The $n$th Fourier coefficient is obtained in
the same manner and is given by equations very similar to
Eqs.~(\ref{eq23} - \ref{eq25}), but involving $I_n(n x)$ and $K_n(n
x)$ instead of $I_1(x)$ and $K_1(x)$.  If this full potential is
used to calculate the intravalley matrix elements, then the $n$th
Fourier term connects the $m=0$ band to the $m=\pm n$ bands.  When
squared and summed to obtain the second order perturbation to the
energy, all of the cross terms vanish and we are left with
\begin{eqnarray}\label{eqa3}
\Delta^{(2)} & = & c_1 a_1^2 + c_2 a_2^2 + c_3 a_3^2 +
\dots \nonumber \\
& \approx & c_1 a_1^2 \left[1 + (a_2/a_1)^2/2 + (a_3/a_1)^2/3 +
\dots\right],\nonumber \\
\end{eqnarray}
where the $c_i$ are constants (independent of the field strength)
and we have approximated $c_n \approx c_1/n$, the factor of $1/n$
coming from the energy denominator.  In Table~\ref{tab2} we list
$a_i/a_1$, for $i = 2 - 4$, for the DNA-CNT system studied in
Sec.~\ref{sec4}. Using $a_2/a_1 \lesssim 0.2$, we find that
including the second Fourier term only changes the second order
energy shift by about $2\%$.  The energy shift was already much less
than the unperturbed energy, and a $2\%$ change to such a small
shift can obviously be neglected.

\begin{table*}
\caption{\label{tab3} Calculated ratios of the $n$th unscreened
Fourier coefficient to the first screened Fourier coefficient for
the nanotube system studied in Sec.~\ref{sec4}.}
\begin{ruledtabular}
\begin{tabular}{cccccccc}
$c$ (nm) & $a_1^{(u)}/a_1^{(s)}$ & $a_2^{(u)}/a_1^{(s)}$ & $a_3^{(u)}/a_1^{(s)}$ & $a_4^{(u)}/a_1^{(s)}$ & $a_5^{(u)}/a_1^{(s)}$ & $a_6^{(u)}/a_1^{(s)}$ & $a_7^{(u)}/a_1^{(s)}$\\
\hline
2        &     $190$             &          $13$         &        $1.1$          &        $0.10$         &  $1.1 \times 10^{-2}$ & $1.2 \times 10^{-3}$  & $1.3 \times 10^{-4}$ \\
5        &     $200$             &          $35$         &        $7.8$          &         $2.0$         &                $0.52$ &               $0.15$  & $4.2 \times 10^{-2}$ \\
10       &     $190$             &          $41$         &         $12$          &         $3.6$         &                 $1.2$ &               $0.42$  &               $0.15$ \\
50       &     $170$             &          $40$         &         $12$          &         $4.0$         &                 $1.5$ &               $0.56$  &               $0.22$ \\
\end{tabular}
\end{ruledtabular}
\end{table*}

\subsection{\label{append01.2} Intervalley Scattering}

The effects of intervalley scattering can be included with an
effective Hamiltonian derived in Ref.~(\onlinecite{an1998}),
\begin{widetext}
\begin{eqnarray}\label{eqa4}
H_{\rm{eff}} = \left(\begin{array}{cccc} u_A(r) & -i\hbar v_f
\left(\partial_x - i \partial_y\right) & -\omega e^{i\theta}
\tilde{u}_A^*(r) & 0 \\ -i\hbar v_f \left(\partial_x +
i\partial_y\right) & u_B(r) & 0 & e^{-i\theta} \tilde{u}_B^*(r) \\
-\omega^* e^{-i\theta} \tilde{u}_A(r) & 0 & u_A(r) & -i\hbar
v_f\left(\partial_x + i\partial_y\right) \\ 0 & e^{i \theta}
\tilde{u}_B(r) & -i\hbar v_f\left(\partial_x - i \partial_y\right) &
u_B(r)\end{array}\right),
\end{eqnarray}
\end{widetext}
where the states are written in the $KA,KB,K^\prime A, K^\prime B$
basis, $\theta$ is the chiral angle, and $\omega = \exp(2\pi i/3)$.
The effective potentials in Eq.~(\ref{eqa4}) are given by
\begin{eqnarray}\label{eqa5}
u_A(\vec{r}) & = & \sum_{R_A} g(\vec{r}-\vec{R}_A)
U(\vec{R}_A),\nonumber\\
u_B(\vec{r}) & = & \sum_{R_B} g(\vec{r}-\vec{R}_B)
U(\vec{R}_B),\nonumber \\
\tilde{u}_A(\vec{r}) & = & \sum_{R_A} g(\vec{r}-\vec{R}_A)
e^{i(\vec{K}-\vec{K}^\prime)\cdot \vec{R}_A} U(\vec{R}_A),\nonumber
\\
\tilde{u}_B(\vec{r}) & = & \sum_{R_B} g(\vec{r}-\vec{R}_B)
e^{i(\vec{K}-\vec{K}^\prime)\cdot \vec{R}_B} U(\vec{R}_B) ,
\end{eqnarray}
where $U(\vec{r})$ is given by Eq.~(\ref{eqa2}), $\vec{R}_A =
\vec{R} + \vec{\tau}_A$ ($\vec{R}_B = \vec{R} + \vec{\tau}_B$)
denotes a site on the $A$ ($B$) sublattice, and $g(\vec{r}-\vec{R})$
is a function peaked at $R$ with a width of about a lattice constant
and normalized so that $\sum_R g(\vec{R}) = 1$. The effective
potentials contain information about the external chiral potential
and the underlying graphene lattice, and in general do not share the
same symmetry as the applied potential. Furthermore, $u_i$ and
$\tilde{u}_i$ do not share the same symmetry, and there is no
coordinate transformation analogous to Eqs.~(\ref{eq02}) and
(\ref{eq03}) to express these potentials in terms of a single
variable.

Second order perturbation theory shows that the off-diagonal matrix
elements contribute in the same way as the diagonal matrix elements
(that is, we have to compute $|\langle K | \tilde{u} | K^\prime
\rangle|^2/(E_K - E_{K^\prime})$). Thus, to get an estimate for the
energy shift due to the off-diagonal terms we need only calculate
the largest Fourier coefficient of those terms.  As mentioned in
Sec.~\ref{sec2}, the Coulomb potential is nearly unscreened at high
momentum transfer, and the unscreened Fourier coefficients of the
potential can be much larger than the screened coefficients.  In
Table~\ref{tab3} we list $a_n^{(u)}/a_1^{(s)}$, the ratio of the
$n$th unscreened Fourier coefficient to the first screened Fourier
coefficient, for the nanotube system studied in Sec.~\ref{sec4}.
From the size of the coefficients it appears intervalley scattering
could be more important than intravalley scattering.

The off-diagonal matrix elements are evaluated as
\begin{equation}\label{eqa6}
\langle K,k_x,m | \tilde{u}_i | K^\prime,k_x^\prime,n \rangle
\propto \sum_{n \neq 0} a_n \sum_G \delta_{\vec{p} - \vec{Q}_n +
\vec{K} - \vec{K}^\prime, \vec{G}},
\end{equation}
where $\vec{p} = (k_x - k_x^\prime)\hat{x} + (m - n + 2\delta)/R
\hspace{1 mm} \hat{y}$, the sum on $n$ is over all integers except
$0$, and the sum on $G$ is over all reciprocal lattice vectors of
the graphene lattice.  This may be rewritten as
\begin{equation}\label{eqa7}
\langle K,k_x,m | \tilde{u}_i | K^\prime,k_x^\prime,n \rangle
\propto \sum_{n \neq 0} a_n \sum_{K_i^\prime} \delta_{\vec{p} -
\vec{Q}_n,\vec{K}_i^\prime},
\end{equation}
where the sum is over all $K^\prime$ points in reciprocal space.
Because of the magnitudes of $\vec{p}$ and $\vec{Q}_n$, only the
$K^\prime$ points on the edge of the first Brillouin may contribute
to this sum.  These points are a function of the chiral angle and
are given by
\begin{eqnarray}\label{eqa8}
\vec{K}_1^\prime & = & \frac{2\pi}{3a}\left\{\left(\cos\theta +
\sqrt{3}\sin\theta\right)\hat{x} + \left(\sqrt{3}
\cos\theta-\sin\theta\right)\hat{y}\right\} \nonumber \\
\vec{K}_2^\prime & = & \frac{2\pi}{3a}\left\{\left(\cos\theta -
\sqrt{3}\sin\theta\right)\hat{x} -\left(\sin\theta + \sqrt{3}
\cos\theta\right)\hat{y}\right\} \nonumber \\
\vec{K}_3^\prime & = & -\frac{4\pi}{3a}\cos\theta \hat{x} +
\frac{4\pi}{3a}\sin\theta\hat{y}.
\end{eqnarray}

\begin{table*}
\caption{\label{tab4} This table shows the $7$ cases where both
inequalities (\ref{eqa9}) and (\ref{eqa10}) are satisfied. The range
of $c$ and $R$ over which the inequalities are satisfied is given,
along with the ratio of the unscreened coefficient to the screened
coefficient and the fractional change in the energy shift. In the
first column $i$ refers to a specific $K_i^\prime$ point.}
\begin{ruledtabular}
\begin{tabular}{ccccccc}
$i$ & $n$  &  $\theta$  & Range of $c$ (nm) & Range of $R$ (nm) & $a_n^u/a_1^s$ & $\left(a_n^u/a_1^s\right)^2/|n|$\\
\hline
$1$ & $3$  &  $\pi/12$  & $1.9 - 2.2$       & $0.30 - 0.35$     & $0.75 - 1.2$  & $0.19 - 0.48$ \\
$1$ & $4$  & $-\pi/12$  & $3 - 13$          & $0.30 - 0.32$     & $0.75 - 1.75$ & $0.14 - 0.76$ \\
$1$ & $4$  &       $0$  & $4 - 5.5$         & $0.30 - 0.33$     & $0.75 - 1.3$  & $0.14 - 0.42$ \\
$1$ & $5$  &  $-\pi/6$  & unrestricted      & $0.37 - 0.38$     & $0.75 - 0.90$ & $0.11 - 0.16$ \\
$1$ & $5$  & $-\pi/12$  & $15 - 20$         & $0.37 - 0.38$     & $0.75 - 0.80$ & $0.11 - 0.13$ \\
$2$ & $-4$ &   $\pi/6$  & $>20$             & $0.30 - 0.31$     & $\sim 1.75$   & $\sim 0.76$   \\
$2$ & $-5$ &   $\pi/6$  & $>15$             & $0.37 - 0.38$     & $0.75 - 0.80$ & $0.11 - 0.13$ \\
\end{tabular}
\end{ruledtabular}
\end{table*}

To remain in the low energy regime we require $|\vec{p}| \lesssim
1/a$, which implies that the only terms contributing to the sum in
Eq.~(\ref{eqa7}) will have
\begin{equation}\label{eqa9}
|\vec{Q}_n - \vec{K}_i^\prime| \lesssim 1/a.
\end{equation}
The expression on the left-hand side of Eq.~(\ref{eqa9}) depends on
$n$, $c$, $R$, $\theta$, and on the particular choice of $K^\prime$
point.  We undertook a systematic search of the relevant parameter
space to determine when Eq.~(\ref{eqa9}) was satisfied.  The search
was limited to $-6 \leq n \leq 6$ because it is clear from
Table~\ref{tab3} that the coefficients of higher Fourier terms will
always be negligible.  Preliminary investigations showed it was
sufficient to restrict the chiral angle to $\theta = 0,\pm \pi/12$
or $\pm \pi/6$.  For each of these $180$ cases we plotted
$|\vec{Q}_n - \vec{K}_i^\prime|$ as a function of $c$ and $R$ to
find where the inequality (\ref{eqa9}) is satisfied.  The search was
restricted to physical values of the pitch and the radius, $1
\hspace{1 mm} \rm{nm} \leq c \leq 60 \hspace{1 mm} \rm{nm}$ and $0.3
\hspace{1 mm} \rm{nm} \leq R \leq 1.0 \hspace{1 mm} \rm{nm}$.  We
found $26$ cases where there was any region in the $(c,R)$ plane
where the inequality was satisfied.

For each of these regions we evaluated the ratio
$a_n^{(u)}/a_1^{(s)}$ to determine if the size of the intervalley
scattering matrix element is comparable to the size of the
intravalley scattering matrix element.  As a conservative estimate
the ratio was considered significant if
\begin{equation}\label{eqa10}
a_n^{(u)}/a_1^{(s)} \geq 0.75.
\end{equation}
There are $7$ cases where the intervalley matrix element is
kinematically allowed (inequality (\ref{eqa9}) is satisfied) and the
unscreened coefficient is large enough to be considered significant.
These $7$ cases are listed in Table~\ref{tab4}, where it is seen
that both inequalities are satisfied only for very small radius
nanotubes, and then only in a narrow range of radii.  Likewise, the
pitch must generally be restricted to a narrow range.  In an
arbitrary sample of nanotubes the fraction of tubes that satisfy
such restrictions will be small, and the intravalley scattering
effects will dominate the intervalley scattering effects in any
measurement.

It may be possible to prepare a sample in one of the allowed regions
in Table~\ref{tab4}, but even then intervalley effects will be no
more important than the intravalley scattering already calculated.
The energy shifts due to intravalley scattering are so small
compared to the unperturbed energy that adding an additional shift
of the same order of magnitude will not make the effects of a chiral
potential any easier to observe. Furthermore, the additional shifts
due to intervalley scattering are small enough that they do not
alter any of the conclusions from the main body of the paper.
Lastly, note that these results were calculated assuming no
screening.  If there is any residual screening either by the
nanotube or by the environment, then the shifts due to intervalley
scattering will be further suppressed.

\bibliography{DNACNTbib}
\end{document}